\documentclass[preprint,aps,draft,showpacs]{revtex4}
\usepackage{graphicx}
\usepackage{epsf}
\tighten
\begin{document}


\title{Triaxiality and the determination of the cubic shape parameter
  K3 from five observables}

\author{V. Werner, C. Scholl, P. von Brentano}

\address{$^1$ Institut f\"ur Kernphysik, Universit\"at zu K\"oln, 
                50937 K\"oln, Germany}

\date{\today}


\begin{abstract}

The absolute and the relative quadrupole shape invariants $q_3$ and
$K_3$ provide a model independent measure of triaxiality for
$\beta$-rigid nuclei. We will show that one can obtain $q_3$ and $K_3$
from a small number of observables. The approximations which are made
will be shown to hold within a few percent both in the rigid triaxial
rotor model and the interacting boson model. The shape parameter $K_3$
is given for an exemplary set of nuclei and is translated into
effective values of the geometrical deformation parameters $\beta$ and
$\gamma$.

\end{abstract}


\maketitle

\section{Introduction \label{intro}}

One basic property of the nucleus is its geometric shape. Therefore,
the nuclear shape, whether it is spherical, prolate, oblate, axially
symmetric, or triaxial, is a key property of the ground state, as well
as of excited states of the nucleus. Quantifying the nuclear shape,
one usually turns to the well known geometric deformation parameters
$\beta$ and $\gamma$. These are deduced from a comparison of data
with, {\it e.g.}, the Davydov-Fillipov model of a rigid triaxial rotor
\cite{DaF58}. This approach incorporates a major problem. A rigid
rotor model cannot account for vibrations of the nuclear shape, which
is a strong limitation. But, even if a model is able to describe also
vibrations in the deformation parameters as, {\it e.g.}, by the Bohr
Hamiltonian \cite{Boh52}, the geometric interpretation of the
interacting boson model \cite{AriIac75,IacAri87}, or the GCM
\cite{Gne69}, a second problem arises. In general the shape parameters
$\beta$ and $\gamma$ do not have fixed values, because the nuclei have
in general not a rigid shape but they are vibrating. Thus it is useful
to consider alternative parameters related to the shape of a nucleus,
namely quadrupole shape invariants \cite{Kum72,Cli86,Jol97}, which are
model independent, and which are direct observables. In this paper we
will discuss mainly the quadratic and cubic shape parameters $q_2$ and
$q_3$. We will focus on the relative cubic shape parameter $K_3 =
q_3/q_2^{3/2}$, which is independent of the nuclear radius $R_0$ and
the charge $e$. We will show that it is possible to obtain $q_2$,
$q_3$ and $K_3$ with good accuracy from only few data. The cubic shape
parameter $K_3$ is related to triaxiality and will be given for a
variety of nuclei. Its connection to the geometrical deformation
parameters will be discussed.

Quadrupole shape invariants were introduced by Kumar \cite{Kum72} and
widely used by Cline and co-workers, {\it e.g.} \cite{Cli86}. They are
expectation values in a given nuclear eigenstate of higher order
moments of the E2 transition operator, which is usually taken to be
the quadrupole operator. Considering the ground state they are defined
as
\begin{eqnarray}
\label{eq:q2def}
q_2 = & & e^2 \langle 0^+_1|(Q\cdot Q)|0^+_1\rangle \ , \\
\label{eq:q3def}
q_3 = & \sqrt{\frac{35}{2}} \  & e^3 \langle
0^+_1|[QQQ]^{(0)}|0^+_1\rangle \ , \\
\label{eq:q4def}
q_4 = & & e^4 \langle 0^+_1|(Q\cdot Q) \ (Q\cdot Q)|0^+_1\rangle \ ,
\end{eqnarray}
where the dot denotes a scalar product and brackets denote tensorial
coupling, $Q$ is the quadrupole operator and $e$ the elementary
electric charge. Higher order moments can also be defined and are
related to fluctuations in $q_3$, but will not be discussed here. The
moments $q_2$ and $q_3$ can be written in terms of averages of
geometrical deformation parameters as
\begin{equation}
\label{eq:q23geo}
q_2 = e^2 Q_0^2 \langle \beta^2\rangle = e^2 Q_0^2 \beta_{eff}^2 \
{\rm and} \ q_3 = e^3 Q_0^3 \langle \beta^3\cos(3\gamma)\rangle \ .
\end{equation}
with 
\begin{equation}
\label{eq:q0}
Q_0 = 3ZR_0^2/(4\pi) \ .
\end{equation}
These quadrupole shape invariants can be renormalized to the second
order invariant $q_2$ by \cite{Jol97,Wer00}
\begin{equation}
\label{eq:kndef}
K_n = \frac{q_n}{{q_2}^{n/2}} \ ,
\end{equation}
omitting the nuclear radius or the electric charge in this form.
These quantities can in principle be obtained directly from data, but
this is difficult in praxis because a large number of E2 matrix
elements including signs is involved in expressions
(\ref{eq:q2def}-\ref{eq:q4def}). This can be seen expanding the
invariants $q_n$ into sums over E2 matrix elements, which is shown
here for $q_2$ and $q_3$:
\begin{eqnarray}
\label{eq:q2sum}
q_2 = & & e^2 \sum_i \langle 0^+_1 \parallel Q\parallel 2^+_i\rangle 
                 \langle 2^+_i \parallel Q\parallel 0^+_1 \rangle \ , \\
\label{eq:q3sum}
q_3 = & \sqrt{\frac{7}{10}} \  & e^3 \sum_{i,j} 
                  \langle 0^+_1 \parallel Q\parallel 2^+_i\rangle 
                  \langle 2^+_i \parallel Q\parallel 2^+_j\rangle 
                  \langle 2^+_j \parallel Q\parallel 0^+_1 \rangle \ .
\end{eqnarray}
An evaluation of $q_2$ and $q_3$ using extensive sets of experimental
quadrupole matrix elements from multiple Coulomb excitation has been
done for some nuclei by D. Cline and co-workers, {\it e.g.} in
\cite{Cli86,Wu96,Wu91}. Of course, the existence of such extensive
data sets is the favorable, but it is not the general case. Thus,
there is great interest to obtain the shape invariants from more
restricted sets of data.

\section{Approximations \label{approx}}

The basic idea is to invoke the $Q$-phonon scheme as has been
discussed in \cite{Wer02}. This scheme was suggested by T. Otsuka 
\cite{Ots94}, and was developed by a K\"oln-Tokyo collaboration, {\it
  e.g.} \cite{Sie94,Pie94,Pie95,Pie98}. The $Q$-phonon scheme implies
that the wave functions of low-lying states are exhausted by only a
few multiple $Q$-phonon configurations, where a $Q$-phonon itself is
an excitation by the quadrupole operator. The $2^+_1$ state in an
even-even nucleus is dominantly a one-$Q$-phonon state. It was shown
\cite{Pie94,Pie95} that the $Q$-phonon scheme holds with good accuracy
for the lowest levels of collective nuclei. Here, as we will consider
only the very lowest states, we keep within the non-orthogonalized
$Q$-phonon scheme \cite{Pie98,Pal98}, which will be shown to be
sufficient for our purpose. The $Q$-phonon scheme gives a simple
selection rule, namely, that an E2 transition between two states may
change the number of $Q$-phonons in first order only by one, {\it
  i.e.} $\Delta Q=1$. Neglecting all $Q$-forbidden transition matrix
elements with $\Delta Q\ge 2$ gives the first order approximation. We
will denote quantities given in this first order approximation by a
superscript $(1)$. This leads to a drastic truncation in the matrix
elements needed in the expansions in
Eqs. (\ref{eq:q2sum},\ref{eq:q3sum}), {\it e.g.}, $q_2$ as given in
Eq. (\ref{eq:q2sum}) is approximated by
\begin{equation}
\label{eq:q2appr}
q_2\approx q_2^{(1)} = e^2 \langle 2^+_1||Q||
0^+_1\rangle^2 = B(E2;0^+_1\rightarrow 2^+_1) \ ,
\end{equation}
because transitions from the two-phonon $2^+_2$ state or even
higher-lying $2^+$ states to the ground state are $Q$-forbidden in
first order. Eq. (\ref{eq:q2appr}) reflects the well-known fact that
in most even-even nuclei the largest part of the E2 excitation
strength is concentrated in the first excited $2^+$ state. In the
rigid rotor this B(E2) value is known to be directly proportional to
the squared $\beta$-deformation by Eq. (\ref{eq:q23geo}). In the
case of non-rigid $\beta$-deformation Eq. (\ref{eq:q23geo}) defines an
effective deformation parameter $\beta_{eff}$ or, making use of the
approximation (\ref{eq:q2appr}), an approximate $\beta_{eff}^{(1)}$.

Using the $Q$-phonon scheme in first order for $q_3$ one obtains
\begin{equation}
\label{eq:q3appbad}
q_3^{(1)} = \sqrt{\frac{7}{10}} \ e^3 \langle 2^+_1||Q||
0^+_1\rangle^2 \cdot \langle 2^+_1||Q|| 2^+_1\rangle \ .
\end{equation}
Then, approximating the $K_3$-parameter following its definition in
Eq. (\ref{eq:kndef}) ($n=3$) results in
\begin{equation}
\label{eq:k3app}
K_3^{(1)} = \frac{q_3^{(1)}}{(q_2^{(1)})^{3/2}} = \sqrt{\frac{7}{10}}
\ \frac{\langle 2^+_1||Q|| 2^+_1\rangle}{\langle 2^+_1||Q||
  0^+_1\rangle} \ ,
\end{equation}
which is calculated from the ratio of the quadrupole moment of the
$2^+_1$ state and its E2 matrix element to the ground state. It turns
out ,{\it e.g.}, checking this approximation within the rigid triaxial
rotor model or the IBM-1, that such a rude truncation of the sum given
in Eq. (\ref{eq:q3sum}) is not sufficient for a good approximation to
$K_3$, as we will show in sections \ref{rotor} and
\ref{ibm}. Therefore, we used a second order approximation, allowing
in each term of the sum {\it only one} $Q$-forbidden matrix element
with $\Delta Q = 2$. Doing so, we derive a second order approximation
for $q_3$ as
\begin{equation}
\label{eq:q3app}
q_3^{(2)} = q_3^{appr.} = \sqrt{\frac{7}{10}} \ e^3 [\langle 2^+_1||Q||
0^+_1\rangle^2 \cdot \langle 2^+_1||Q|| 2^+_1\rangle + 2 \cdot \langle
0^+_1||Q|| 2^+_2\rangle \cdot \langle 2^+_2||Q|| 2^+_1\rangle \cdot
\langle 2^+_1||Q|| 0^+_1\rangle] \ ,
\end{equation}
which still includes only few E2 matrix elements. In the following we
will always denote the second order approximation with the superscript
$appr.$ instead of $(2)$, as it is the only one we use. Note that the
approximation to $q_2$ in second order approximation is the same as in
first order, as a $Q$-forbidden matrix element would always appear
squared and such terms are not included in this approximation, and we
get
\begin{equation}
\label{eq:q2app2}
q_2^{appr.} = q_2^{(1)} \ {\rm and} \ \beta_{eff}^{appr.} =
\beta_{eff}^{(1)} \ .
\end{equation}
Dividing $q_3^{appr.}$ from Eq. (\ref{eq:q3app}) by $q_2^{appr.}$, we
get a second order approximation for $K_3$ that includes only four
different E2 matrix elements, involving the lowest two excited $2^+$
states and the ground state.

A problem that appears is that the signs of the E2 matrix elements are
needed, which are not known in most cases. Usually we know only the
B(E2) values which are
\begin{equation}
\label{eq:be2}
B(E2;J_i\rightarrow J_f) = \frac{1}{2J_i +1} \ e^2 \langle
J_f||Q||J_i\rangle^2 \ .
\end{equation}
This ambiguity in the signs can be avoided by using a relation between
the signs of four matrix elements, suggested by Jolos and von Brentano
\cite{Jol96}:
\begin{equation}
\label{eq:phasrel}
{\rm sign}( \langle 2^+_1||Q|| 2^+_1\rangle ) = - {\rm sign}( \langle
0^+_1||Q|| 2^+_2\rangle \langle 2^+_2||Q|| 2^+_1\rangle \langle
2^+_1||Q|| 0^+_1\rangle ) \ .
\end{equation}
This relation gives the relative phase of the two terms in
Eq. (\ref{eq:q3app}). There is still an overall sign of $K_3$, which
is the sign of the quadrupole moment of the $2^+_1$ state, deciding
between prolate and oblate deformation. Then, the second order
approximation for $K_3^{appr.}$ is
\begin{equation}
\label{eq:k3appsig}
K_3^{appr.} = \sqrt{\frac{7}{10}} \ {\rm sign}(Q(2^+_1)) \ \left[
\sqrt{\frac{B(E2;2^+_1\rightarrow 2^+_1)}{B(E2;2^+_1\rightarrow
0^+_1)}} - \frac{\sqrt{B(E2;2^+_2\rightarrow 0^+_1) \cdot
B(E2;2^+_2\rightarrow 2^+_1)}}{B(E2;2^+_1\rightarrow 0^+_1)} \right] \
,
\end{equation}
where we use an alternative but elegant definition of the squared
quadrupole moment following Eq. (\ref{eq:be2}), 
\begin{equation}
\label{eq:be2q}
B(E2;2^+_1\rightarrow 2^+_1) = \frac{1}{5} \ e^2 \langle 2^+_1||Q||
2^+_1\rangle^2 = \frac{35}{32\pi} \ Q(2^+_1)^2 \ .
\end{equation}

The approximation formula for $K_3$ given in Eq. (\ref{eq:k3appsig})
is the key result of this work. It allows to measure this observable 
directly and in a model independent way from only few data. These are
four absolute B(E2) values, namely $B(E2;2^+_1\rightarrow 0^+_1)$,
$B(E2;2^+_2\rightarrow 0^+_1)$, $B(E2;2^+_2\rightarrow 2^+_1)$, and
$B(E2;2^+_1\rightarrow 2^+_1)$, and the sign of the quadrupole moment of
the $2^+_1$ state, which we consider as a fifth observable. This new
method to determine $K_3$ is of particular importance because $K_3$ is
closely connected to the triaxiality of nuclei, {\it i.e.} to 
$\gamma$-deformation. For axial symmetry $K_3=-1$ for prolate
($\gamma=0^{\circ}$) and $K_3=+1$ for oblate ($\gamma=60^{\circ}$)
nuclei, while $K_3$ drops to zero at a maximum triaxiality of
$\gamma=30^{\circ}$. This holds for geometrical models like the
Davydov-Fillipov triaxial rotor model, as well as for the dynamical
symmetries of the IBM. One major difference between these two models
is that the IBM describes non-rigid $\beta$- and $\gamma$-deformation,
{\it e.g.}, in the U(5) vibrational limit and the O(6) limit of
$\gamma$-soft nuclei, in both of which $K_3$ vanishes. In the SU(3)
and $\overline{{\rm SU(3)}}$ dynamical symmetries of the IBM, which
correspond to the prolate and oblate axially symmetric rigid rotors,
respectively, the same values for $K_3$ are derived as in the
geometrical model. In the following we will check to which extent 
\begin{equation}
\label{eq:k3k3app}
K_3\approx K_3^{appr.}
\end{equation}
holds, using as a test the rigid triaxial rotor model of Davydov and
Fillipov and the IBM-1.

\section{The rigid triaxial rotor model \label{rotor}}

The Hamiltonian of the Davydov-Fillipov rotor model is
\begin{equation}
\label{eq:hdav}
H_{geo} = A_1 J_1^2 + A_2 J_2^2 + A_3 J_3^2 \ ,
\end{equation}
where $J_n$ are the projections of the spin $J$ on the three symmetry
axes, and where the parameters $A_k$ are connected to the moments of
inertia $\Theta_k$ by
\begin{equation}
\label{eq:ak}
A_k = \frac{\hbar}{2\Theta_k} \ .
\end{equation}
The moments of inertia can further be written in terms of the
geometrical deformation parameters $\beta$ and $\gamma$,
\begin{equation}
\label{eq:momi}
\Theta_k = 4 B \beta^2 \sin^2\left(\gamma-\frac{2k}{3}\pi \right) \ .
\end{equation}
The E2 transition operator is given by
\begin{equation}
\label{eq:qgeo}
T(E2)_{geo} = e Q_{2\mu} = e Q_0 \ \beta \ \left[
D^{2\ast}_{\mu 0} \cos(\gamma) + \sqrt{2} \ (D^{2\ast}_{\mu 2} +
D^{2\ast}_{\mu -2}) \sin(\gamma) \right] \ ,
\end{equation}
where the $D^2_{\mu\nu}$ are the Wigner-$D$-matrices and $Q_0$ is
given by Eq. (\ref{eq:q0}). We stress that this model with rigid
$\beta$- and $\gamma$-deformations is applicable to only a limited
number of nuclei. Nevertheless, apart from our discussion of the
ground state deformation, the model is often applied to highly excited
and strongly and also super-deformed bands as well, for which, in
principle, our approach of $K$-parameters may also apply.

\begin{figure}[hbt]
\epsfxsize 8cm
  \epsfbox{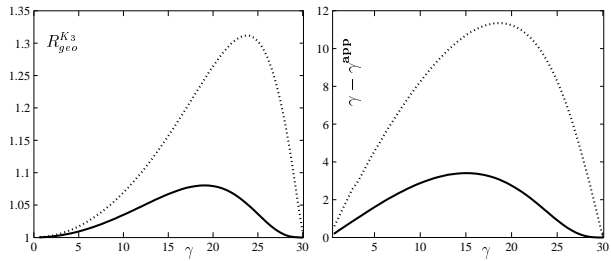}
\caption{$R_{geo}^{K_3}$ calculated for all values of $\gamma$ (solid
  line on the left hand side), showing that the second order
  approximation Eq. (\ref{eq:k3k3app}) holds well in the rigid
  triaxial rotor model. The approximation does not seriously change
  the value of $\gamma$ (solid line on the right hand side). The
  dashed lines give the values derived from the use of only the first
  order approximation.}
\label{fig:k3geo} 
\end{figure}

In our calculations we vary the parameter $\gamma$ over the range
($\gamma\in [0^{\circ},30^{\circ}]$), covering the range of prolate
axially symmetric and triaxial structures inherent to the
model. The results for $\gamma\in [30^{\circ},60^{\circ}]$ are fully
symmetric to those given and thus omitted. The choice of $\beta$ is
arbitrary, as in the rigid case $K_3$ is independent of $\beta$ and
is given by 
\begin{equation}
\label{eq:k3geo}
K_3 = - \frac{\beta^3\cos(3\gamma)}{(\beta^2)^{3/2}} = -\cos(3\gamma) \ .
\end{equation}
In a similar way one defines an approximate deformation
$\gamma^{appr.}$ from $K_3^{appr.}$ by
\begin{equation}
\label{eq:gapp}
K_3^{appr.} = -\cos(3\gamma^{appr.}) \ .
\end{equation}
In order to avoid the division by zero, we use the ratio
\begin{equation}
\label{eq:rgeo}
R_{geo}^{K_3} = \frac{1+|K_3^{appr.}|}{1+|K_3|}
\end{equation}
as a measure of the quality of the approximation
(\ref{eq:k3k3app}). The solid curve in the left panel of Figure
\ref{fig:k3geo} shows the quantity $R_{geo}^{K_3}$ versus the
deformation parameter $\gamma$, calculated numerically using the code
{\sc Davidov} \cite{davi}. In the axially symmetric limit at
$\gamma=0^{\circ}$ the approximation is exact. This also holds for the
case of maximum triaxiality at $\gamma=30^{\circ}$, while
$R_{geo}^{K_3}$ is small for all intermediate cases with a deviation
from one of 8$\%$ in maximum. The dashed curve represents the same
calculation, but using the first order approximation $K_3^{(1)}$ from
Eq. (\ref{eq:k3app}). The deviation of the first order approximation
is clearly much larger with a maximum of about 30$\%$, showing that
the use of a second order approximation is unavoidable for
transitional nuclei. On the right hand side of Figure \ref{fig:k3geo}
the corresponding absolute deviation of $\gamma^{appr.}$ derived from
Eq. (\ref{eq:gapp}) from the real $\gamma$-values in the model is
shown as a solid curve. The maximum deviation is below 3.5$^{\circ}$
at $\gamma\approx 15^{\circ}$. Again, the deviation is much larger
using only the first approximation, given as a dashed curve..

\section{The interacting boson model \label{ibm}}

\begin{figure}[hbt]
\epsfxsize 8cm
  \epsfbox{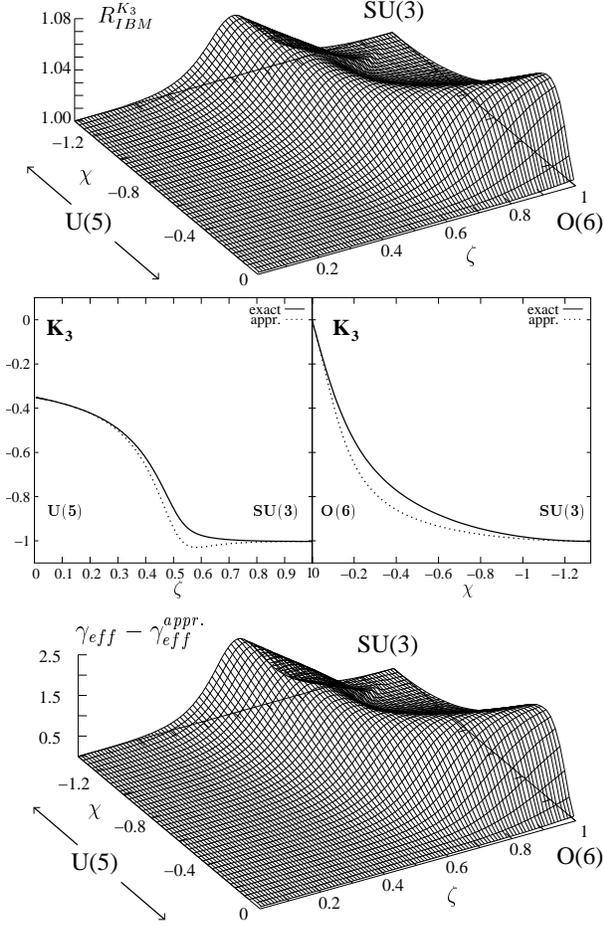}
\caption{Top panel: $R_{IBM}^{K_3}$ calculated over the whole IBM
  parameter space confirms a good fulfillment of the second order
  approximation to $K_3$. Middle panel: The two transitional legs for
  fixed values of $\chi=-\sqrt{7}/2$ (left) and $\zeta=1$ (right). The
  approximation misses slightly the phase-/shape-transitional parameter
  region. Bottom panel: Effective $\gamma$-deformations calculated from
  $K_3$ and $K_3^{appr.}$ calculated over the whole parameter range. All
  calculations are for $N=10$ bosons.}
\label{fig:k3ibm} 
\end{figure}

\begin{figure}[hbt]
\epsfxsize 8cm
  \epsfbox{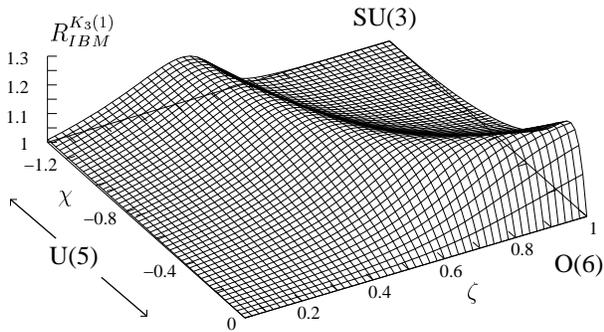}
\caption{$R_{IBM}^{K_3 (1)}$, the analog to the top panel of
  Figure \ref{fig:k3ibm}, but using only the first order
  approximation. Deviations from unity are much larger than in the
  second approximation.}
\label{fig:k3ibmbad} 
\end{figure}

Now, we check the quality of $K_3^{appr.}$ in the IBM-1, within the
Extended Consistent Q Formalism (ECQF) \cite{WarCas82,Lip85} and the
Hamiltonian \cite{Wer00}
\begin{equation}
\label{eq:hecqf}
H_{IBM} = (1-\zeta) \ n_d - \frac{\zeta}{4N} \ Q^{\chi} \cdot Q^{\chi}
\ ,
\end{equation}
depending on only two structural parameters, $\zeta$ and $\chi$, and
omitting an overall energy scale. The E2 transition operator in the
ECQF is chosen to be proportional to the quadrupole operator in the
Hamiltonian,
\begin{equation}
\label{eq:qibm}
T(E2)_{IBM} = e_B Q^{\chi} = e_B [(s^+{\tilde d} + d^+s) + \chi
(d^+{\tilde d})] \ ,
\end{equation}
with an effective boson charge $e_B$, and $n_d=(d^+{\tilde d})$ is the
boson number operator. Varying the values of $\zeta$ and $\chi$ over
the full range of symmetries ($\zeta\in [0,1],
\chi\in[-\sqrt{7}/2,\sqrt{7}/2]$), one covers the dynamical symmetry
limits of the IBM, namely U(5) ($\zeta=0,\chi$), the prolate (oblate)
SU(3) ($\overline{\rm SU(3)}$) ($\zeta=1,\chi=\mp\sqrt{7}/2$), and
O(6) ($\zeta=1,\chi=0$), as well as the transitional structures in
between. In analogy to Eq. (\ref{eq:rgeo}), we define the ratio
\begin{equation}
\label{eq:ribm}
R_{IBM}^{K_3} = \frac{1+|K_3^{appr.}|}{1+|K_3|} \ ,
\end{equation}
which has been calculated over the full parameter space using the code
{\sc Phint} \cite{phint}. Again, $K_3^{appr.}$ is defined by
Eq. (\ref{eq:k3appsig}). The results are shown in the top part of Figure
\ref{fig:k3ibm} for $N=10$ bosons. They are given for $\chi<0$,
because the results for $\pm\chi$ are fully symmetric, as the change
in sign is equivalent to the symmetry transformation $d\rightarrow
-d$ (keeping $s\rightarrow s$). The use of positive $\chi$ values
corresponds to the choice of $\gamma > 30^{\circ}$ in the geometrical
model. Deviations of $K_3^{appr.}$ from the exact $K_3$ values are
small in all cases, the deviation of $R_{IBM}^{K_3}$ from 1 is below
7$\%$. For comparison, Figure \ref{fig:k3ibmbad} shows $R_{IBM}^{K_3
  (1)}$, which is defined analog to Eq. (\ref{eq:ribm}), but where the
first order approximation, {\it i.e.} Eq. (\ref{eq:k3app}), is
used. Like in the geometrical model it is seen, that the deviations
from the exact value of $K_3$ are much larger in the first order
approximation. Thus, in general it is necessary to use $K_3^{appr.}$
from Eq. (\ref{eq:k3appsig}).

The deviation $R_{IBM}^{K_3}$ peaks in a region around SU(3). A
reason for this behavior is found by a close look at this
region. The middle part of Figure \ref{fig:k3ibm} shows the values of
$K_3$ and $K_3^{appr.}$ on the U(5)--SU(3) (left) and O(6)--SU(3)
(right) transition legs. It is obvious that the maximum deviation of
$R_{IBM}^{K_3}$ from unity appears in those regions, in which $K_3$
changes most rapidly. These are exactly those regions that are
connected to the shape/phase transition between spherical and axially
symmetric nuclei, or between prolate and oblate deformations, as
discussed, {\it e.g.}, in
\cite{Iac00,Cas00,Iac01,Cas01,Wer02,JJ01}. This means that in the IBM 
the approximation $K_3^{appr.}$ misses the exact value of $K_3$
somewhat when leaving the rotational limit. However, overall
deviations of $R_{IBM}^{K_3}$ from unity are small and the approximation
(\ref{eq:k3k3app}) is well fulfilled.

From comparison with the geometrical model an effective
$\gamma$-deformation can be defined \cite{Wer00} from $K_3$ by
\begin{equation}
\label{eq:geff}
K_3^ = - \frac{\langle \beta^3\cos(3\gamma)\rangle}{\langle
\beta^2\rangle^{3/2}} = - \cos(3\gamma_{eff}) \ ,
\end{equation}
and an approximate value $\gamma_{eff}^{appr.}$ can be defined analog
from $K_3^{appr.}$. The differences between the exact and the
approximate $\gamma$-values, $\gamma_{eff}-\gamma_{eff}^{appr.}$ are
included in the bottom part of Figure \ref{fig:k3ibm} and show good
agreement. The deviation of $\gamma_{eff}^{appr.}$ from $\gamma_{eff}$
is always smaller than 2.5$^{\circ}$.  

These effective $\gamma$-values are not and cannot be equivalent to
those given by Eqs. (\ref{eq:k3geo},\ref{eq:gapp}), because $K_3$ is
not generally independent of $\beta$-deformation and fluctuations in
$\beta$ occur, especially for vibrational nuclei. Moreover, in case of
rigid $\beta$ (on the SU(3)--O(6) transitional line) $K_3$ is a
measure of $\langle \cos(3\gamma)\rangle$, while in case of rigid
$\gamma$ it is a measure of $\langle \beta^3\rangle/\langle
\beta^2\rangle^{3/2}$. The effect of a $\beta$-vibration is only
effectively taken out in the translation to the geometric model by
Eq. (\ref{eq:geff}). However, if fluctuations in $\beta$ are small,
which is the case past the phase transition towards deformed nuclei
(typically for $\zeta > 0.6$), a factorization of the averages over
$\beta$ and $\cos(3\gamma)$ should work, and we can assume
\begin{equation}
\label{eq:average}
\langle \beta^3\cos(3\gamma)\rangle = \langle \beta^3\rangle \langle
\cos(3\gamma)\rangle \ {\rm and} \ \frac{\langle \beta^3\rangle}{\langle
\beta^2\rangle^{3/2}} = 1 \ ,
\end{equation}
making $\gamma_{eff}$ comparable to the geometrical $\gamma$-deformation.

\section{$K_3$ for various nuclei \label{data}}

\subsection{Direct measure of $K_3$ \label{direct}}

For the two considered models we have shown that $K_3^{appr.}$ is a
good approximation to the value of the cubic shape parameter
$K_3$. Thus we assume this to hold also in other collective models
such as the GCM or the Bohr Hamiltonian. Only few observables have
to be obtained in order to derive $K_3^{appr.}$, namely the lifetime
of the $2^+_1$ state, the lifetime and the branching ratio of the
$2^+_2$ state, and the quadrupole moment of the $2^+_1$ state. Besides
the modulus of $K_3$ also its sign is interesting, which is obtained
from the sign of the quadrupole moment of the $2^+_1$ state. This
quadrupole moment itself is not easy to obtain, therefore it is a
challenge to measure triaxiality. Especially for vibrational or
$\gamma$-soft nuclei, where the quadrupole moment is small, high
quality data is needed. Thus, an approximate value of $K_3$ is so far
known for a number of nuclei in or near the valley of stability
only. For a set of nuclei that belong to various symmetry regions the
$K_3^{appr.}$-parameter has been calculated from tabulated data. The
results are given in Table \ref{tab:k3nuc}, together with effective
$\gamma$-deformation parameters derived from Eq. (\ref{eq:geff}), and
effective $\beta$-deformations from
Eqs. (\ref{eq:q23geo},\ref{eq:q2appr}).

Typical rotational nuclei like the heavier Gd or
Dy isotopes show $K_3$ values around -1 as it is expected for prolate
deformed axially symmetric shapes. Also $^{152}$Sm and $^{154}$Gd,
which are attributed \cite{Cas00,Ton04} to be close to the critical
point symmetry X(5) proposed by F. Iachello \cite{Iac00}, show this
value. The $K_4$-parameter obtained from $q_4$ of Eq. (\ref{eq:q4def})
and Eq. (\ref{eq:kndef}), which can be approximated in a similar way
\cite{Jol96,Wer02}, gives a direct measure for $\beta$-softness. One
finds that $K_4^{appr.}=1$ for $\beta$-rigid nuclei and
$K_4^{appr.}\sim 1.4$ for vibrators. For $^{152}$Sm and $^{154}$Gd one
finds $K_4^{appr.}$ values of 1.02(3) \cite{Klu00} and 1.088(26)
\cite{Ton04}, respectively, which, in combination with $K_3^{appr.}$,
meets the expectations for the vibrator to well-deformed rotor
transition, more on the rotational side of the phase
transition. Especially $^{152}$Sm seems to be on the rotor side (where
$K_3^{appr.}=-1$ and $K_4^{appr.}=1$) of the phase transition, which
seemingly conflicts with the interpretation of this nucleus as being
close to the phase transitional point. This may be related to the
systematical error made in the approximations for $K_3$, which
maximizes exactly in the transitional region in the IBM, which may be
reflected also in data. However, the systematical error made in the
determination of $K_4$ should be smaller in that region \cite{Jol96}
and a problem remains.

The $K_3$ values of the Os isotopes show an evolution from the axially
symmetric rotor towards O(6) symmetry with a maximum effective
triaxiality of $\gamma_{eff}=30^{\circ}$. Note, that here one talks of
{\it effective} $\gamma$-deformation, as the nucleus does not have a
rigid triaxiality. The more vibrational Pd and Cd isotopes show
moderate values of $K_3$ with relatively large errors due to the
quadrupole moments. Non-zero values are not a contradiction to a more
U(5) like structure as they may emanate from finite N effects (see
\cite{Wer00}).

An surprising conflict appears for $^{196}$Pt, which is usually
taken as a prime example of O(6) symmetry \cite{Ciz78}, as well as for
the neighboring $^{194}$Pt. Both nuclei show rather large, positive
quadrupole moments \cite{Wu96} and thus have quite large values of
$K_3^{appr.}$. This shows that they are on the side of oblate
deformation, with a considerable deviation of $K_3^{appr.}$ from the 
expectation value, $K_3(O(6))=0$. Other observables like the branching
ratio of the $2^+_2$ state or energies agree much better with the O(6)
predictions. Therefore, values of $K_3$ derived from an IBM fit (see
below) agree much better with $K_3=0$. One cannot argue that this
deviation is due to the dependence on $\beta$-fluctuations (compare
Eq. (\ref{eq:geff})). On the SU(3)--O(6) transition line, no
$\beta$-fluctuations are allowed, and indeed, the shape invariant
$K_4$ approximately equals 1 (see \cite{Wer02}) for both nuclei, which
pinpoints $\beta$-rigidity. Again, this may be related
to the maximal systematical error close to O(6) seen from Figure
\ref{fig:k3ibm}. But, even if the value of $K_3$ is overpredicted from
the approximation, a deviation from O(6) remains. However, we want to
stress that these values, {\it e.g.} $\gamma_{eff}=42^{\circ}$ instead
of $\gamma_{eff}=30^{\circ}$ for $^{196}Pt$, still indicate a strong
triaxiality. It is only the quantitative value of $\gamma_{eff}$ which
is in doubt.

We stress that if one uses only the first order approximation, the
value of $K_3$ is missed for transitional nuclei like the Os isotopes,
for which the transition $2^+_2\rightarrow 0^+_1$ is sizeable, {\it
  e.g.}, $K_3^{appr.} = -0.7(1)$ for $^{190}$Os in the first order
approximation, underestimating triaxiality, while the second order
approximation gives $K_3^{appr.} = -0.35(9)$.
 
\begin{figure}[hbt]
\epsfxsize 8cm
  \epsfbox{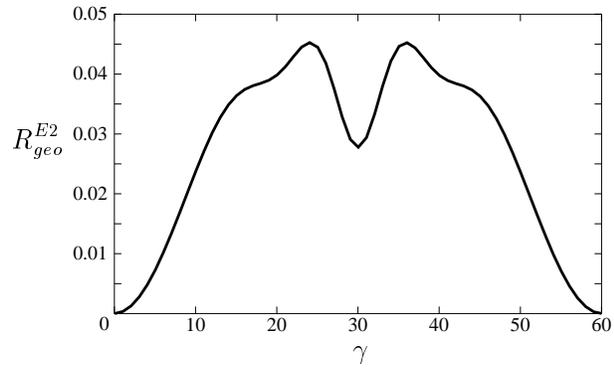}
\caption{$R_{geo}^{E2}$ calculated for all values of $\gamma$. The
E2-relation (\ref{eq:be2dav}) holds well in the geometrical model.}
\label{fig:be2dav} 
\end{figure}

Note, that for the electric quadrupole moment of the $2^+_1$ state,
$Q(2^+_1)$, is usually not easy to access experimentally. A new relation
was found \cite{Wer02}, however, which gives a new way to
approximately determine $Q(2^+_1)$ or $B(E2;2^+_1\rightarrow 2^+_1)$,
respectively:
\begin{equation}
\label{eq:be2rel}
B(E2;2^+_1\rightarrow 2^+_1)^{appr.} = B(E2;4^+_1\rightarrow 2^+_1) -
B(E2;2^+_2\rightarrow 2^+_1) \ .
\end{equation}
This relation may be used to obtain the quadrupole moment as an input
for $K_3$, but it will give a large uncertainty especially for
vibrational or $\gamma$-soft nuclei, which have a small quadrupole
moment. So far the relation (\ref{eq:be2rel}) was only checked in the
IBM \cite{Wer02}. Figure \ref{fig:be2dav} shows the deviation 
\begin{equation}
\label{eq:be2dav}
R_{geo}^{E2} = 1 - \frac{B(E2;4^+_1\rightarrow
  2^+_1)}{B(E2;2^+_1\rightarrow 2^+_1) + B(E2;2^+_2\rightarrow
  2^+_1)}
\end{equation}
from the real value calculated within the rigid triaxial rotor
model. Also in the geometrical model the agreement is good.

\subsection{Fit procedure for $K_3$ \label{fit}}

In cases where not all of the needed data are present, one may follow
another procedure, fitting parameters of a model to the available data
for one nucleus, and calculating $K_3$ from the model. Here we used the
simple two parameter Hamiltonian of the IBM given in
Eq. (\ref{eq:hecqf}). The two parameters were fitted to the
energy ratio 
\begin{equation}
\label{eq:r42}
R_{4/2}=E(4^+_1)/E(2^+_1)
\end{equation}
and the B(E2) ratio
\begin{equation}
\label{eq:b22}
B(E2;2^+_2\rightarrow 0^+_1)/B(E2;2^+_2\rightarrow 2^+_1) \ ,
\end{equation}
that are sensitive to changes in structure over wide parameter
regions. For the reproduction of the energy ratio an error of 2$\%$
was allowed, while for the B(E2) ratio the experimental errors were
taken into account, resulting in an allowed parameter range of $\zeta$
and $\chi$, in which $K_3$ takes various values within a certain
range, with an upper a lower limit. In Table \ref{tab:k3nuc} we denote
values of $K_3$ obtained from the fit as $K_3^{fit}$, and give the
upper and lower limits allowed from the experimental errors. These
values can be compared with the measured $K_3^{appr.}$. The values
agree reasonably well in most cases, considering the simplicity of the
Hamiltonian and the arbitrary choice of the two observables used in
the fit. Note, that other observables can be used for the fit. But in
some cases the simple Hamiltonian used cannot describe all features of a
given nucleus, as, {\it e.g.}, for the Pt nuclei. Therefore, the
Hamiltonian (\ref{eq:hecqf}) may be extended or another model used.

\section{Conclusions \label{conc}}

To conclude, we discussed measures of triaxiality. In this respect we
considered in particular the absolute and relative cubic shape
parameters $q_3$ and $K_3$. The approximative $K_3^{appr.}$ was
introduced as a direct, model independent observable, which is, if
$\beta$ is rigid, a measure of triaxiality, while $K_3^{appr.}$ is
more general an observable in all structural limits and the regions
between them. $K_3^{appr.}$ was shown to be a good approximation to
the exact value of $K_3$, and can with good accuracy be obtained from
only four matrix elements, or from four B(E2) values, one of them
equivalent to the modulus of the quadrupole moment $Q(2^+_1)$, and the
sign of $Q(2^+_1)$. This, manifested in Eq. (\ref{eq:k3appsig}), is
the key result of this work. This accuracy of the approximation was
checked for the IBM and the rigid triaxial rotor model, and the need
of the second order approximation within the $Q$-phonon scheme was
shown in both models, especially in transitional regions. Effective
values of $\beta$- and $\gamma$-deformation in the ground state,
derived from $q_2^{appr.}$ and $K_3^{appr.}$, respectively, have been
deduced from data. For vibrational nuclei geometrical deformation
parameters cannot be given, while $q_2$ and $K_3$ are always
well-defined properties of the ground state. Finally, we proposed a
way how to derive $K_3^{appr.}$ from a model fitting data.

For discussions we thank A. Dewald, A. Gelberg, J. Jolie, R.V. Jolos,
B.R. Mottelson, and I. Stefanescu. 

This work was supported by the DFG under contract No. Br 799/12-1.

\begin{table}[htb]
\caption{Approximate $K_3^{appr.}$-values, the effective approximate
  $\beta$- and $\gamma$-deformation parameters derived from our
  approach are listed for a set of nuclei. For $\beta$-deformations,
  errors are omitted as they are in the order of per mil or smaller,
  and the systematic error made by assuming $R_0=1.2 fm$ for the
  nuclear radius is presumably larger. The last two columns
  give upper and lower limits for the value of $K_3$ fitted to the
  observables (\ref{eq:r42},\ref{eq:b22}) as described in section
  \ref{fit}.}
\label{tab:k3nuc}
\begin{tabular}{c@{\quad}|@{\quad}c@{\quad}c@{\quad}c@{\quad}cc}
 & $K_{3}^{appr.}$ & $\beta_{eff}^{appr.}$ & $\gamma_{eff}^{appr.}$ &
 \multicolumn{2}{c}{$K_3^{fit}$} \\
 & & & & upper & lower \\
\hline
$^{156}$Gd & -0.97(5) & 0.339 & 4(4)  & -0.86 & -0.98 \\
$^{158}$Gd & -0.95(6) & 0.349 & 6(6)  & -0.83 & -1.00 \\
$^{160}$Gd & -0.96(3) & 0.351 & 5(3)  & -0.85 & -1.00 \\
$^{164}$Dy & -0.93(9) & 0.347 & 7(7)  & -0.77 & -1.00 \\
$^{154}$Gd & -1.00(3) & 0.310 & 1(5)  & -0.77 & -0.84 \\
$^{152}$Sm & -0.98(4) & 0.307 & 4(4)  & -0.54 & -0.83 \\
$^{188}$Os & -0.63(5) & 0.185 & 17(3) & -0.65 & -0.76 \\
$^{190}$Os & -0.35(9) & 0.177 & 23(3) & -0.53 & -0.75 \\
$^{192}$Os & -0.3(1)  & 0.167 & 25(2) & -0.48 & -0.71 \\
$^{194}$Pt & 0.53(4)  & 0.143 & 41(1) &  0.08 &  0.14 \\
$^{196}$Pt & 0.59(7)  & 0.129 & 42(2) &  0.00 &  0.02 \\
$^{106}$Pd & -0.4(1)  & 0.230 & 22(3) & -0.41 & -0.55 \\
$^{108}$Pd & -0.6(4)  & 0.242 & 17(16)& -0.23 & -0.30 \\
$^{112}$Cd & -0.4(1)  & 0.181 & 22(2) & -0.57 & -0.83 \\
$^{114}$Cd & -0.4(1)  & 0.184 & 23(3) & -0.36 & -0.59 
\end{tabular}
\end{table}

\end{document}